\documentclass[aps,preprint,showpacs,superscriptaddress]{revtex4}
\usepackage{amssymb,bm}
\usepackage{graphicx}
\usepackage{amsmath}
\usepackage{epstopdf}
\usepackage{comment}
\allowdisplaybreaks
\usepackage[utf8]{inputenc}
\graphicspath{{Figs/}}
\renewcommand{\Re}{\operatorname{Re}}
\renewcommand{\Im}{\operatorname{Im}}
\usepackage{color}

\begin{document}

\title{High-energy bremsstrahlung on atoms  in  a laser field}
\author{P. A. Krachkov}
\email{P.A.Krachkov@inp.nsk.su}
\affiliation{Max-Planck-Institut f\"ur Kernphysik, Saupfercheckweg 1, 69117 Heidelberg, Germany}
\affiliation{Budker Institute of Nuclear Physics of SB RAS, 630090 Novosibirsk, Russia}
\author{A. Di Piazza}
\email{dipiazza@mpi-hd.mpg.de}
\affiliation{Max-Planck-Institut f\"ur Kernphysik, Saupfercheckweg 1, 69117 Heidelberg, Germany}
\author{A. I. Milstein}
\email{milstein@inp.nsk.su}
\affiliation{Budker Institute of Nuclear Physics of SB RAS, 630090 Novosibirsk, Russia}

\date{\today}

\begin{abstract}
	The impact of a laser field on the process of photon radiation by an ultra-relativistic electron in an atomic field is investigated. The angular distribution and the spectrum of the radiated photon are derived. By means of the quasiclassical approximation, the obtained results are exact in the parameters of the laser field and the atomic field. It is shown that the impact of the laser field is significant even for fairly average values of the laser field parameters routinely achievable nowadays. Therefore, an experimental observation of the influence of the laser field on bremsstrahlung in the atomic field is a very feasible task.
\end{abstract}


\maketitle

\section{Introduction}

Quantum electrodynamics (QED) processes in atomic fields are of great interest from an experimental point of view, because the principles of detection of charged particles and photons at high energies are based on such processes. For example, bremsstrahlung and photoproduction of electron-positron pairs are the main processes determining the propagation of electromagnetic showers in matter. Also, they represent background processes in the study of strong and electroweak interactions, as well as in the search for new physics. For QED processes in an atomic field, the effective coupling constant for heavy atoms is $ \eta = Z \alpha$, which can be of the order of unity. Here, $ Z $ is the atomic charge number, $ \alpha = e ^ 2$ is the fine structure constant, with $e$ being the electron charge, and units with $ \hbar = c = 1 $ are employed. Therefore, in these cases it is necessary to perform calculations exactly in $ \eta $, which is a nontrivial task.

The method of quasiclassical Green's functions of the Dirac equation in atomic fields developed in recent years has made a breakthrough in the theoretical description of the fundamental high-energy QED processes in these fields \cite{KLM16}. The quasiclassical approach allows one to obtain results for an arbitrary atomic potential, taking into account the finite size of the nucleus and screening effects, and without requiring the analytical solution of the Dirac equation.

Another important example of processes in external fields are QED processes in strong laser fields. The recent rapid development of laser technologies makes it possible to produce high-intensity laser fields (intensities up to $ 10^{22} $ W/cm$^2 $), which opens up new opportunities for experimental and theoretical studies of QED in the nonlinear strong-field regime \cite{Mitter_1975,Ritus_1985,Baier_b_1998,Ehlotzky_2009,DiPiazza2012,King_2016}. 

The influence of the laser field on QED processes is characterized by two dimensionless parameters, $ \xi = | e | E / m \omega_0 $ and $ \chi = (\varepsilon / m) (E / E_c) $, where $ m $ is the electron mass, $ E $ and $ \omega_0 $ are the laser electric field amplitude and its angular frequency, 
$ \varepsilon $ is the energy of the incident particle and $ E_c = m^2 / | e | = 1.3 \times 10^ {16} \, \text {V/cm} $  is the critical electric field. Here, we have implicitly assumed that the laser field can be approximated as a plane wave, that the particle is initially counterpropagating with respect to the laser field and that, in the case of massive particles like electrons or positrons, it is ultrarelativistic. Several QED processes in a strong laser field, such as electron radiation in a laser wave, pair production, photon splitting, and others, are being extensively studied and we refer the reader to the reviews \cite{Ritus_1985, Baier_b_1998,Ehlotzky_2009, Di_Piazza_2012,King_2016}.

It is interesting to investigate how the presence of a laser field affects QED processes in a strong atomic field. This question has been studied in detail exactly in the parameters $\xi$ and $ \chi$ but in the leading approximation in $\eta$. For example, the process of electron-positron pair production has been considered in Ref. \cite {Loetstedt_2008, Di_Piazza_2009}, bremsstrahlung in Ref. \cite {Loetstedt_2007}, and Delbr\"uck scattering in Ref. \cite{Di_Piazza_2008}. Calculating the probabilities of such processes with exact account of the parameters of both laser and atomic fields is not an easy task, since there is no exact solutions of the Dirac equation in the superposition of a laser field (even under the plane-wave approximation) and an atomic field. However, the quasiclassical approach allows one to find the Green's function and the wave function of an ultrarelativistic electron in the superposition of an atomic field and a laser field exactly in the parameters of both the external fields, but approximately in the parameter $ m / \varepsilon$ (we recall that $\varepsilon$ is the energy of the incoming particle). Within the quasiclassical approach, the investigation of $e^+e^-$ pair photoproduction in the superposition of atomic and laser fields has been performed in Refs. \cite {DiPiazza2012, DiPiazza2014}. In these works, it was shown that the presence of the laser field induces a suppression of the cross section of $e^+e^-$ pair photoproduction in an atomic field. This effect is similar to the Landau-Pomeranchuk-Migdal effect (LPM) \cite {LanPom, Migdal}, which is the suppression of $e^+e^-$  photoproduction cross section and the bremsstrahlung spectrum at high energies due to multiple scattering by atoms in matter. In the case of electron-positron photoproduction, in order to observe the LPM effect in matter photon energies as high as $ \omega \gtrsim 2.5\,\text{TeV} $ are necessary, which makes extremely hard the experimental observation of the LPM in this case. To observe the effect of suppression of the $e^+e^-$ photoproduction cross section due to a laser field, one can also have much lower photon energies ($ \omega \gtrsim10\,\text{ GeV}$ ) and a laser field strength already available ($ 10^{21}\,W /\text{cm}^2$).

The LPM effect for bremsstrahlung is much easier to observe than a similar effect for photoproduction. In fact, the formation length $l$ for both processes is given by the same formula $ l \sim \lambda_c {\varepsilon \varepsilon '}/ {\omega m} $, where $\omega$ is the photon energy, $ \lambda_c=1/m=3.9 \times 10^{-11}\;\text{cm}$ is the Compton wavelength, and where for photoproduction $ \varepsilon $  and $ \varepsilon '$ are  the electron and positron energies, respectively, whereas for bremsstrahlung $ \varepsilon $  and $ \varepsilon '$ are the initial electron and final electron energies, respectively. Thus, unlike that for photoproduction, in the case of bremsstrahlung the quantity ${\varepsilon'}/{\omega} $ becomes much larger than unity in the soft part of the spectrum $\omega\ll \varepsilon$, resulting in an increase of the formation length $l$. Therefore, multiple scattering has a stronger effect on bremsstrahlung than on photoproduction. This explains why the LPM effect for bremsstrahlung has been already measured \cite{MPL_meas1, MPL_meas2} for incoming electron with energy  $ 25\,\text{GeV} $ in the detected photon energy region $ \omega = 0.5-500\,\text{MeV} $.

In this paper, we study the effect of the laser field on the process of bremsstrahlung of high-energy electrons in an atomic field. Calculations are performed exactly in the parameters $\xi$, $\chi$, and $\eta$, but in the leading approximation in the parameters $ m /\varepsilon \ll 1$ and $ m \xi/\varepsilon \ll 1$. We show that the laser field greatly modifies the bremsstrahlung spectrum and the angular distribution of the outgoing photon. As in the case of the LPM effect in matter, experimental observation of the effect of the laser field on the process of bremsstrahlung is much more favorable than that on the photoproduction process for the same physical reason mentioned above in the case of the LPM effect in matter.

The paper is organized as follows. In Sec. \ref{sec2} we present the derivation of the matrix element of the process. In Sec. \ref{sec3}, we discuss the angular distribution and the spectrum of the emitted photon. In Sec. \ref{sec4}, we consider in detail the case of monochromatic circularly-polarized plane wave. In Sec. \ref{sec5}, we investigate the case of a weak laser field. Finally, in Sec. \ref{sec6}, the main conclusions of the paper are
presented.

\section{Matrix element}\label{sec2}

Let an ultrarelativistic electron with momentum $\bm p$, directed almost along the $z$-axis, and energy $\varepsilon=\sqrt{\bm{p}^2+m^2}$ interact with the atomic potential $V(\bm{r})$ in the presence of a counter-propagating plane wave, described by the vector potential $\bm{A}(t+z)$, with $\bm{z}\cdot\bm{A}(t+z)=0$. We use the quasiclassical electron wave functions,  obtained exactly in the parameters of the atomic and laser fields but in the leading approximation in the parameters $ m /\varepsilon $, $ m \xi/\varepsilon $ and in the angles between the momenta of the final particles and the momentum of the initial electron \cite{DiPiazza2014}.

In the absence of the atomic field, the radiation process in the laser field is described in terms of a probability per unit time. This probability is well known \cite{Ritus_1985,Di_Piazza_2012} and can be subtracted in the final answer. After such a subtraction is carried out, one can describe the radiation process in terms of a cross section. At high energies this cross section reads
 \begin{equation}\label{eq:cs}
 d\sigma=\frac{\alpha}{(2\pi)^4}|M|^{2}\frac{d\omega}{\omega}\,d{\bm{k}}_\perp\,d{\bm{p'}}_\perp\,\,,
 \end{equation}
where $\bm k$ and $\bm p'$ are the photon and the final electron momenta, respectively, $\omega=|\bm k|=\varepsilon-\varepsilon'$ and $\varepsilon'=\sqrt{ \bm p'^2+m^2}$ are the photon and the final photon energies, respectively, and in general $\bm X_\perp$ denotes a component of the vector $\bm X$  perpendicular to $z$-axis. Strictly speaking, different quantities are conserved in the atomic field and in the laser field. In the atomic field energy is conserved, whereas in the laser field (plane wave) the conserved quantities are the transverse momentum $\bm p_\perp$ and the light-cone energy $ p_- = p_z + \varepsilon $. In the ultrarelativistic case and in the counterpropagating setup, it is $ p_- = 2 \varepsilon + O ({m^2} /{\varepsilon^ 2}) $, such that within our approximations the energy is conserved in the superposition of these two fields.

The matrix element  $M$ in Eq. \eqref{eq:cs} has the form
\begin{equation}\label{M12}
M=-\int dT d\bm \rho\,\,\bar U_{p',\mu'}^{(\text{out})}(T,\bm{\rho}) \,\bm e^*\cdot\bm{\gamma}\, e^{-i \bm k_\perp \cdot\bm \rho+i k_\perp^2 T/2\omega }\,\,U_{p,\mu}^{(\text{in})}(T,\bm{\rho})\,,
\end{equation}
where   $\gamma^\nu$  are the Dirac matrices,  $\bm e$ is the emitted photon polarization vector, $T=(t+z)/2$, and $\bm{\rho}=(x,y)$. The wave functions $U_{p,\mu}^{(\text{in})}(T,\bm{\rho})$  and $U_{p',\mu'}^{(\text{out})}(T,\bm{\rho})$ are the solutions  of the Dirac equation in the superposition of the laser and atomic fields, with $\mu$ and $\mu'$ indicating the signs of the electron helicity \cite{DiPiazza2014}. The superscripts $(\text{in})$ and $(\text{out})$ indicate that the asymptotic forms of $U_{p,\mu}^{(\text{in})}(T,\bm{\rho})$ and $U_{p',\mu'}^{(\text{out})}(T,\bm{\rho})$ at large $\bm r$ contain, in addition to the plane wave, the spherical divergent and convergent waves, respectively (we assume that $|\bm{A}(x)|\rightarrow 0$ for $|x|\rightarrow\infty$, without requiring the plane wave to be monochromatic).

The wave functions $U_{p,\mu}^{(\text{in})}(T,\bm{\rho})$ and $\bar U_{p,\mu}^{(\text{out})}(T,\bm{\rho})$ have the form \cite{DiPiazza2014}:

\begin{align}
&U_{p,\mu}^{(\text{in})}(T,\bm{\rho})=\exp\left\{-i\frac{m^2}{2\epsilon}T+i\bm{p}_{\perp}\cdot\bm{\rho}-\frac{i}{2\epsilon}\int_{0}^Td\tau\,[\bm{p}_{\perp}-\bm{\mathcal{A}}(\tau)]^2\right\}\nonumber\\
&\times\left[1-\frac{i}{2\epsilon}\bm{\alpha}\cdot\partial_{\bm{\rho}}-\frac{1-\alpha^3}{4\epsilon}\,\bm{\alpha}\cdot\bm{\bm{\mathcal{A}}}(T)\right]u_{p,\mu}\int \frac{d\bm{q_-}}{i\pi} \exp\left[iq_-^2-i\int_0^{\infty}d\tau\,V(\bm{\rho}_-,T-\tau) \right]\,,\nonumber\\
&\bar U_{p,\mu}^{(\text{out})}(T,\bm{\rho})=\exp\left\{i\frac{m^2}{2\epsilon}T-i\bm{p}_{\perp}\cdot\bm{\rho}+\frac{i}{2\epsilon}\int_0^{T}d\tau[\bm{p}_{\perp}-\bm{\mathcal{A}}(\tau)]^2\right\}\nonumber\\
&\times \bar u_{p,\mu}\left[1-\frac{i}{2\epsilon}\bm{\alpha}\cdot\partial_{\bm{\rho}}+\frac{1-\alpha^3}{4\epsilon}\,\bm{\alpha}\cdot\bm{\bm{\mathcal{A}}}(T)\right]\int \frac{d\bm{q_+}}{i\pi} \exp\left[iq_+^2-i\int_0^{\infty}d\tau\,V(\bm{\rho}_+,T+\tau) \right],\nonumber\\
\bm{\rho}_{\pm}&=\bm{\rho}\pm\tau\frac{\bm{p}_{\perp}}{\epsilon}+\left[\sqrt{\frac{\mp 2T}{\epsilon}}\bm{q_\pm}+\frac{1}{\epsilon}\int_0^Tdy\bm{\bm{\mathcal{A}}}(y)\right]\theta(\mp T)\,,
\end{align}
where $\bm{\mathcal{A}}(T)=e\bm A(2 T)=e\bm A(t+z)$, where $\bm q_\pm$ are two two-dimensional vectors perpendicular to the $z$-axis, where $\bm\alpha=\gamma_0\bm{\gamma}$, and where $u_{p,\mu}$ is the corresponding solution of the free Dirac equation.

Within our accuracy, we write the matrix element $M$ as follows
\begin{align}\label{Matr_str}
M=\bar u_{p',\mu'}\left\{\hat e^*M_0 +\left[\frac{(1-\alpha^3)\bm\alpha \hat e^*}{4\varepsilon'}-\frac{\hat e^*(1-\alpha^3)\bm\alpha}{4\varepsilon}\right]\cdot\bm M_1-\frac{\bm\alpha \hat e^*}{2\varepsilon'}\cdot\bm M_2-\frac{\hat e^*\bm\alpha}{2\varepsilon}\cdot\bm M_3 \right\}u_{p,\mu}\,,
\end{align}
where the quantities $M_0$ and $\bm M_{1,2,3}$ are some functions reported below. For  definite helicities of the particles, the matrix element $M$ reads (cf. the corresponding result in Ref. \cite{KM2015}):
\begin{eqnarray}\label{Mres1}
&&M=\frac{\delta_{\mu\mu'}\bm e^*_\lambda}{\varepsilon\varepsilon'}\cdot\left[\varepsilon\delta_{\lambda\mu}(-\varepsilon'\bm\theta_{p'k}M_0-\bm M_2+\bm M_1)+ \varepsilon'\delta_{\lambda\bar\mu}(-\varepsilon\bm\theta_{pk}M_0+\bm M_3+\bm M_1)\right]
\nonumber\\
&&-\frac{m\mu\delta_{\mu'\bar\mu}\delta_{\lambda\mu}\omega}{\sqrt{2}\varepsilon\varepsilon'}M_0\,,
\end{eqnarray}
where $\lambda$ is the sign of the helicity of the photon and where $\bm\theta_{p}=\bm p_\perp/\varepsilon$, $\bm\theta_{p'}=\bm p'_\perp/\varepsilon'$, $\bm\theta_k=\bm k_{\perp}/\omega$ and $\bm\theta_{xy}=\bm\theta_x-\bm\theta_y$. 
 
First, we consider the term $ M_0 $, which we represent as a sum  $ M_0 = M_0^++ M_0 ^- $, where the terms $ M_0^+ $ and $ M_0^- $ correspond to the contributions of the integral  over positive values of $ T $ and negative values of $ T $, respectively. In the term $M_0^+$  we make the substitution 
$$\bm\rho\to\bm\rho-\sqrt{\frac{ 2T}{\varepsilon}}\bm{q_-}-\frac{1}{\varepsilon}\int_0^Tdy\bm{\bm{\mathcal{A}}}(y)$$
and then take the integrals over $\bm q_+$ and $\bm q_-$. We obtain 
\begin{align}
&M_0^+=\int_{0}^\infty dT\int d\bm \rho\,\exp\Big\{-i \bm \Delta_\perp \cdot\bm \rho-i \Delta_\parallel T-i \Delta_\perp^2 T/2\varepsilon\nonumber\\ &+i\frac{\omega}{2\varepsilon\varepsilon'}\int_0^Td\tau [\bm{\mathcal{A}}^2(\tau) -2\varepsilon'\bm{\mathcal{A}}(\tau)\cdot\bm\theta _{p'k}]\nonumber\\
&-i\int_0^{\infty}d\tau\,[V(\bm{\rho}+\tau\bm \theta_{p'},T+\tau)+\,V(\bm{\rho}-\tau\bm \theta_{p},T-\tau)]\Big\},
\end{align}
where $\bm\Delta=\bm{p}'+\bm{k}-\bm{p}$. Similarly, we obtain for the term $ M_0^- $
\begin{align}
&M_0^-=\int_{-\infty}^0 dT\int d\bm \rho\,\exp\Big\{-i \bm \Delta_\perp \cdot\bm \rho-i \Delta_\parallel T+i \Delta_\perp^2 T/2\varepsilon'\nonumber\\ &+i\frac{\omega}{2\varepsilon\varepsilon'}\int_0^Td\tau [\bm{\mathcal{A}}^2(\tau) -2\varepsilon\bm{\mathcal{A}}(\tau)\cdot\bm\theta _{pk}]\nonumber\\
&-i\int_0^{\infty}d\tau\,[V(\bm{\rho}+\tau\bm \theta_{p'},T+\tau)+\,V(\bm{\rho}-\tau\bm \theta_{p},T-\tau)]\Big\}\,.
\end{align}

The influence of the laser field on the bremsstrahlung cross section is most important at high electron energies since the parameter $\chi=(\varepsilon/m)(E/E_c)$ is proportional to $\varepsilon/m$. Already at relatively low energies $\varepsilon\gtrsim 100\,\text{MeV}$ the formation length $l=\varepsilon \varepsilon'/\omega m^2$ is significantly larger than the screening radius $r_{scr}\sim Z^{-1/3}/m\alpha$ (the so-called case of full screening). It is this energy region that we consider in our work.
In the expressions for the amplitudes in this energy region, all angles in the integrand  are  small. Using this  circumstance we can make in the term $M_0^+$ the substitutions $T\to T-\bm\rho\cdot\bm\theta_{p}$,
$\bm\rho\to \bm\rho+T\bm\theta_{p}$ and the replacement  $V(\bm{\rho}+\tau\bm \theta_{p'},T+\tau)\rightarrow V(\bm{\rho}+\tau\bm \theta_{p},T+\tau)$. Similarly, in  $M_0^-$ these transformations are $T\to T-\bm\rho\cdot\bm\theta_{p'}$, $\bm\rho\to \bm\rho+T\bm\theta_{p'}$, and $V(\bm{\rho}-\tau\bm \theta_{p},T-\tau)\rightarrow V(\bm{\rho}-\tau\bm \theta_{p'},T-\tau)$. As a result we have
\begin{align}
&M_0=\Xi(\bm \Delta_\perp) [\Psi_+(\varepsilon'\bm\theta_{p'k})+ \Psi_-(\varepsilon\bm\theta_{pk})]\,,\nonumber\\
&\Xi(\bm{\Delta_\perp})=\int d\bm\rho\,\exp[-i\bm\Delta_\perp\cdot\bm\rho-i\mathcal{V}(\bm\rho)]\,,\quad\mathcal{V}(\bm\rho)=\int_{-\infty}^{\infty}dz\,V(\rho,z)\,,\nonumber\\
&\Psi_\pm(\bm X)=\int_0^\infty dT\,\, \exp\left\{i\frac{\omega}{2\varepsilon\varepsilon'}\int_0^{\pm T}d\tau [(\bm X-\bm{\mathcal{A}}(\tau))^2+m^2]\right\}\,.
\end{align}

The remaining terms are obtained in the same way and the result is
\begin{align}\label{M_int}
&\bm M_1= \Xi(\bm \Delta_\perp)\left[\bm \Psi_+^{{\mathcal{A}}}(\varepsilon'\bm\theta_{p'k})+ \bm\Psi_-^{{\mathcal{A}}}(\varepsilon\bm\theta_{pk})\right]\,,\nonumber\\
&\bm M_2=-\bm\Delta_{\perp}\Xi(\bm \Delta_\perp)\Psi_-(\varepsilon\bm\theta_{pk})\,,\quad
\bm M_3=-\bm\Delta_{\perp}\Xi(\bm \Delta_\perp)\Psi_+(\varepsilon'\bm\theta_{p'k})\,,\nonumber\\
&\bm\Psi_\pm^{\mathcal{A}}(\bm X)=\int_0^\infty dT\,\, \bm{\mathcal{A}}(\pm T) \exp\left\{i\frac{\omega}{2\varepsilon\varepsilon'}\int_0^{\pm T}d\tau [(\bm X-\bm{\mathcal{A}}(\tau))^2+m^2]\right\}\,.
\end{align}

By substituting Eq.~\eqref{M_int} in Eq.~\eqref{Mres1}, by summing over the helicities of the final particles and averaging over the initial electron helicity, we find
\begin{align}\label{Mres}
&\overline{M^2}=\frac{R(\bm{\Delta_\perp})}{2\varepsilon^2\varepsilon'^2}\left[(\varepsilon^2+\varepsilon'^2)|\bm f_1+\bm g_1|^2+m^2\omega^2 |f_0+ g_0|^2\right]\,,\nonumber\\
&R(\bm{\Delta_\perp})=\int d\bm \rho_1 d\bm \rho_2\,e^{-i\bm\Delta_\perp\cdot(\bm\rho_1-\bm\rho_2)} \left[ e^{-i\mathcal{V}(\bm\rho_1)+i\mathcal{V}(\bm\rho_2)}-1\right]\,,\nonumber\\
&f_0=\Psi_+(\varepsilon'\bm\theta_{p'k})\,,\quad g_0=\Psi_-(\varepsilon\bm\theta_{pk})\,,\nonumber\\
&\bm f_1=\bm\Psi_+^{{\mathcal{A}}}(\varepsilon'\bm\theta_{p'k})-\varepsilon'\bm\theta_{p'k}\Psi_+(\varepsilon'\bm\theta_{p'k})\,,\nonumber\\
&\bm g_1= \bm\Psi_-^{{\mathcal{A}}}(\varepsilon\bm\theta_{pk})-\varepsilon\bm\theta_{pk}\Psi_-(\varepsilon\bm\theta_{pk})\,.
\end{align}
In Eq. \eqref{Mres} we have subtracted a contribution independent of the atomic field (corresponding to the term $-1$ in the square bracket in the expression of $R(\bm{\Delta_\perp})$). The expression \eqref{Mres} together with Eq. \eqref{eq:cs} defines the differential bremsstrahlung cross section in the superposition of the atomic and laser fields. By integrating over $\bm p_\perp'$ one obtains the angular distribution of photons at fixed $\omega$. Then, taking the integral over $\bm k_\perp$ one obtains the expression for the photon spectrum.

\section{Angular distribution and spectrum of photons}\label{sec3}
In order to derive the angular distribution of the process, we substitute Eq. \eqref {Mres} in Eq.\eqref {eq:cs}, pass from the variable $ \bm p_\perp' $ to $ \varepsilon' \bm \theta_{p'k} $ and take the integrals over $\varepsilon'\bm\theta_{p'k} $, $ \bm\rho_1 $ and 
$ \bm\rho_2 $, using the relation
\begin{align}\label{CC1}
\int d\bm \rho&\left[ e^{-i\mathcal{V}(\bm\rho+\bm x)+i\mathcal{V}(\bm\rho-\bm x)}-1\right]=-8\pi\eta^2x^2\left[L-\log{mx}-C\right]\,,
\end{align} 
valid for full screening within the Thomas-Fermi model \cite{LM95A} when $\log{mx}\ll L$.
Here
\begin{align}
L= \log183 Z^{-1/3}-\mbox{Re}\,\psi(1+i\eta)-C\,,
\end{align} 
where $C$ is the Euler constant and $\psi(x)=d\ln\Gamma(x)/dx$.
Then, the differential cross section can be written as the sum of two terms:
\begin{align}\label{ang_distr}
&\frac{d\sigma}{d\omega d \bm k_\perp}=\frac{d\sigma_1}{d\omega d \bm k_\perp}+\frac{d\sigma_2}{d\omega d \bm k_\perp}\,,\nonumber\\
&\frac{d\sigma_1}{d\omega d \bm k_\perp}=-\frac{4\alpha\eta^2}{\pi m^4\omega D} \Im \int d \tau_1 d \tau_2 \tau_1 e^{i\phi_+}\Bigg\{ \left[2L-C+i\dfrac{\pi}{2}-\log \tau_1+i\zeta_1 F(\zeta_1)\right]\nonumber\\
&\times\left[\left(1 +i\zeta_1\right) (\bm v \cdot \bm v_-+D)- \bm v_-\cdot\bm b_1
\right]\nonumber\\
&-\bm v_-\cdot(\bm v -\bm b_1) -D 
-\left(1-e^{-i\zeta_1}\right)\left[D + \bm v \cdot\bm v_-+i\frac{\bm v_-\cdot\bm b_1}{\zeta_1}
\right]
\Bigg\}\,,\nonumber\\
&\frac{d\sigma_2}{d\omega d \bm k_\perp}=-i\frac{2\alpha\eta^2}{\pi m^4\omega D}\int d \tau_1 d \tau_2 e^{i\phi_-}\Bigg\{ \left[2L-C+i\dfrac{\pi}{2}-\log \tau_-+i\zeta_2 F(\zeta_2)\right]\nonumber\\
&\times\left[i -\tau_-(1+i\zeta_2)(D+\bm v\cdot\bm v_+) +\tau_-\bm b_2 \cdot(\bm v + \bm v_+) \right]\nonumber\\
&-2i +\tau_-(\bm v -\bm b_2)\cdot(\bm v_+-\bm b_2)-\zeta_2 +\tau_-D \nonumber\\
&+\tau_-\left (1-e^{-i\zeta_2}\right)\left[i\frac{(\bm v +\bm v_+)\cdot\bm b_2}{\zeta_2}
+\bm v \cdot\bm v_+ +D\right]
\Bigg\}\,,\nonumber\\
&\tau_\pm=\tau_1\pm\tau_2\,,\quad\bm b=\frac{\varepsilon\bm{\theta}_{pk}}{m}\,,\quad \bm b_1=\bm b-\int_0^{\tau_1}\frac{\bm{\mathcal{ A}}_x }{\tau_1} dx\,,\quad \bm b_2=\bm b-\int_{\tau_2}^{\tau_1}\frac{\bm{\mathcal{ A}}_\tau}{\tau_-} d\tau\,, \nonumber\\
& \bm v_\pm= \bm{\mathcal{ A}}_{\pm\tau_2}-\bm b\,,\quad \bm v =\bm{\mathcal{ A}}_{\tau_1}-\bm b\,,\quad\bm{\mathcal{ A}}_\tau=\frac{1}{m}\bm{\mathcal{A}}\Bigr(\frac{2\varepsilon \varepsilon'\tau}{\omega m^2}\Bigr)\,,\quad
\zeta_1=\tau_1 b_1^2\,,\quad\zeta_2=\tau_- b_2^2\,,\nonumber\\
&\phi_\pm=\tau_\pm+\int_{\mp\tau_2}^{\tau_1}(\bm{\mathcal{ A}}_\tau-\bm b)^2 d\tau\,,\quad
F(x)=\int_0^1 dt e^{-ixt}\log t\,,\quad D=\frac{\omega^2}{\varepsilon^2+\varepsilon'^2} \,.
\end{align}
Note that the contribution $ d \sigma_1/d \omega d\bm k_\perp $ is given by the terms  in Eq. \eqref{Mres} proportional to $ f_0 g_0^* $ and to $ \bm f_1 \cdot\bm g_1^* $ , whereas the contribution $ d \sigma_2/d \omega d \bm k_\perp $ by the terms proportional to $ | f_0 |^2 $ and to $ | \bm f_1 |^2 $. The terms  $ | g_0 |^ 2 $ and $ | \bm g_1 | ^ 2 $ do not contribute to $ d \sigma/d \omega \bm d k_\perp $. Using Eq. \eqref{ang_distr} it is possible to obtain the cross section $d\sigma_{pp}/d\varepsilon\,d\bm p_\perp$ of $e^+e^-$ pair production by a photon with the energy $\omega$ in combined atomic and laser fields, where $\varepsilon$ is the produced electron energy, $\bm p$ is the electron momentum, and $\bm p_\perp$ is the transverse component of $\bm p$ with respect to $\bm k$. In order to do this it is necessary to perform the substitutions $\bm p \rightarrow -\bm p$, $\varepsilon \rightarrow -\varepsilon$, $\bm k \rightarrow -\bm k$, and $\omega \rightarrow -\omega$.

Taking the integral over $\bm k_{\perp}$ in Eq. \eqref{ang_distr} we arrive at the photon spectrum 
\begin{align}\label{spectr}
&\frac{d\sigma}{d\omega}=-\frac{4\alpha\eta^2\omega}{m^2\varepsilon^2D}\Re \int \frac{d \tau_1 d \tau_2}{\tau_+^2} e^{i(\tau_++\phi)}\Bigg\{\left[2 L-C+i\dfrac{\pi}{2}-\log\frac{\tau_1 \tau_2}{\tau_+}+i\zeta F(\zeta)\right]\nonumber\\
&\times\left[\frac{2i\tau_1\tau_2}{\tau_+}+i\zeta \bm{s}_1\cdot\bm{s}_2- (\bm{\beta}+\bm{s}_1)\cdot(\bm{\beta}-\bm{s}_2)+D \tau_1\tau_2 (1+i\zeta) \right]\nonumber\\
&-\frac{3i\tau_1\tau_2}{\tau_+}+ (\bm{\beta}+\bm{s}_1)\cdot(\bm{\beta}-\bm{s}_2)-D \tau_1\tau_2 -\frac{\zeta\tau_1\tau_2}{\tau_+}\nonumber\\
&-\left(1-e^{-i\zeta}\right)\left[D\tau_1\tau_2
+\frac{i\tau_1\tau_2}{\tau_+}+i\frac{(\bm{s}_1-\bm{s}_2)\cdot\bm{\beta}}{\zeta}+ \bm{s}_1\cdot\bm{s}_2 
\right]
\Bigg\}\,,\nonumber\\
&\phi=\int_{-\tau_2}^{\tau_1}\bm{\mathcal{A_\tau}}^2 d\tau-\frac{(\int_{-\tau_2}^{\tau_1}\bm{\mathcal{A_\tau}} d\tau)^2}{\tau_+} \,,\quad \zeta=\bm{\beta}^2\frac{\tau_+}{\tau_1\tau_2}\,,\quad\bm{\beta}=\frac{\tau_2}{\tau_+}\int_0^{\tau_1}\bm{\mathcal{ A}}_\tau d\tau+\frac{\tau_1}{\tau_+}\int_0^{-\tau_2}\bm{\mathcal{ A}}_\tau d\tau\,,\nonumber\\
&\bm{s}_1 =\tau_1 \bm{\mathcal{ A}}_{\tau_1}-\frac{\tau_1}{\tau_+}\int_{-\tau_2}^{\tau_1}\bm{\mathcal{ A}}_\tau d\tau\,,
\quad \bm{s}_2 =\tau_2 \bm{\mathcal{ A}}_{-\tau_2}-\frac{\tau_2}{\tau_+}\int_{-\tau_2}^{\tau_1}\bm{\mathcal{ A}}_\tau d\tau\,.
\end{align}

Note that the term $d\sigma_2/d\omega d \bm k_\perp$ does not contribute to the spectrum, because ${d\sigma}/{d\omega}$ is determined by the interference of the terms in Eq.~\eqref{M12} corresponding to the integration over positive and negative values of $T$ ($ f_0$ and $g_0$, $ \bm f_1$  and $\bm g_1 $, respectively). Using Eq. \eqref{spectr} it is possible to obtain the cross section $d\sigma_{pp}/d\varepsilon$ by multiplying Eq. \eqref{spectr} by the factor $\varepsilon^2/\omega^2$ and by performing the substitution $\varepsilon \rightarrow -\varepsilon$ and $\omega \rightarrow -\omega$.

The expressions \eqref{ang_distr} and  \eqref{spectr} are obtained for an arbitrary phase dependence of the potential $\bm{\mathcal{ A}}(T)$ and any values of the laser field parameters. Below we consider in detail some special cases of $\bm{\mathcal{ A}}(T)$. 

\section{Monochromatic circularly-polarized plane wave}\label{sec4}
For a circularly-polarized  monochromatic plane wave we have
\begin{align}
&\bm{\mathcal{ A}}_\tau=\xi\, [\cos(\Omega\tau)\,\bm e_1+\sin(\Omega\tau)\,\bm e_2] \,,\quad \Omega=\frac{4\varepsilon \varepsilon'\omega_0}{\omega m^2}\,,
\end{align}
where  $\bm e_1$ and $\bm e_2$ are two unit vectors perpendicular to $z$-axis, $\omega_0$ it the laser angular frequency, $\xi=| e | E / m \omega_0 $, and  $ E $ is the amplitude of the electric field of the laser. In this case the expression of the spectrum considerably simplifies and it reads
\begin{align}\label{spectrmono}
&\frac{d\sigma}{d\omega}=-\frac{4\alpha\eta^2\omega}{m^2\varepsilon^2D}\Re \int \frac{d \tau_1 d \tau_2}{\tau_+^2} e^{i\tau_+[1+\varkappa(\tau_+)]}\Bigg\{\left[2 L-C+i\dfrac{\pi}{2}-\log\frac{\tau_1 \tau_2}{\tau_+}+i\zeta F(\zeta)\right]\nonumber\\
&\times\left[\frac{\tau_1\tau_2}{\tau_+}(2i-\zeta)-G_2+G_1 (1+i\zeta) \right]-\frac{3i\tau_1\tau_2}{\tau_+}+ G_2 -G_1-\left(1-e^{-i\zeta}\right)\left[G_1
+\frac{i\tau_1\tau_2}{\tau_+}+i\frac{G_2}{\zeta} 
\right]
\Bigg\}\,,\nonumber\\
&G_1=-\tau_1\tau_2\left[\varkappa(\tau_+)+2f(\tau_+)-2\xi^2\cos^2(\tau_+\Omega/2)-D\right]\,,\quad f(\tau)=\xi^2 \frac{\sin(\Omega\tau)}{\Omega\tau}\,, \nonumber\\
&G_2=\left[f(\tau_1)-f(\tau_+)+\frac{\tau_-}{2\tau_+}\left( \varkappa(\tau_1)-\varkappa(\tau_+)\right)\right]+(\tau_1\leftrightarrow\tau_2)\,,\quad \varkappa(\tau)=\xi^2\left[1-\frac{\sin^2(\Omega\tau/2)}{(\Omega\tau/2)^2}\right]\,, \nonumber\\
&\quad\zeta=\tau_+\varkappa(\tau_+)-\tau_1\varkappa(\tau_1)-\tau_2\varkappa(\tau_2)\,,
\end{align}
where $D$ is defined in Eq. \eqref{ang_distr}.
If one passes in Eq. \eqref{spectrmono} from the variables $\tau_1$ and $\tau_2$ to the variables
$\tau=\tau_+$ and $y=\tau_1/\tau_+$, one finds that for $\Omega=1+\xi^2$ there is a divergence of the integral at large $\tau$. The condition $\Omega=1+\xi^2$ is equivalent to the relation
\begin{equation}
\omega=\omega^*=\dfrac{4\varepsilon^2 \omega_0}{m^2(1+\xi^2)+4\varepsilon \omega_0}\,.
\end{equation}
Now, for optical lasers ($\omega_0\sim 1\,\mbox{eV}$) we have  $\varepsilon\omega_0/m^2\ll 1$ for electron energies up to $\varepsilon\sim {20}\,\mbox{GeV}$, and below for simplicity we write all expressions in the leading  approximation in the parameter $\varepsilon\omega_0/m^2$. Thus,
\begin{equation}
\omega^*=\dfrac{4\varepsilon^2 \omega_0}{m^2(1+\xi^2)}\ll \varepsilon\,.
\end{equation}
The quantity $\omega^*$ is nothing but the maximum frequency of the photon produced in a collision of the electron with the energy $\varepsilon $ and a circularly-polarized plane wave with the frequency  $ \omega_0 $. The divergence is related to the cascade process when an electron radiates  a photon in a pure laser field and then scatters in the atomic field (or vice versa). Exactly at the resonance ($\Omega=1+\xi^2$) the expression \eqref{spectrmono} is not applicable because the relation \eqref{CC1} used in our derivation is valid under the condition $\log{mx}\ll L\sim \log(1/mr_{scr})$. As a consequence of this condition, we find that  Eq. \eqref{spectrmono} is applicable if $|\omega/\omega^*-1|(1+\xi^2)\gg 1/(mr_{scr})^2$. This statement follows from the asymptotics of  Eq. \eqref{spectrmono} obtained at $|\omega/\omega^*-1|\ll \xi$:
\begin{align}\label{spectr_res}
&\dfrac{d\sigma}{d\omega}=\frac{4\alpha\eta^2\xi^2\omega^*}{3 m^2 (1+\xi^2)^2(\omega^*-\omega)^2}
\left\{L+\frac{1}{2}\log\left[\frac{|\omega-\omega^*|(1+\xi^2)}{\omega^*}\right]-\frac{1}{6}\right\}\,.
\end{align}
We emphasize that, for the case of a circularly-polarized laser field, due to the conservation of the $z$-component of the total angular momentum the absorption of more than one laser photon is strongly suppressed in the ultrarelativistic regime under investigation. The absorption of two laser photons, in particular, can occur only if the electron helicity changes sign in the emission process and the emission probability with spin flip  is $m^2/\varepsilon^2$ times smaller than the probability of emission without spin flip.

For $\omega\ll \omega^*$  and any value of $\xi$, we obtain 
\begin{align}
\frac{d\sigma}{d\omega}=&\frac{16\eta^2\alpha}{3m^2(1+\xi^2)\omega}\left[L+\frac{1}{2}\log (1+\xi^2)+\frac{1}{12} \right]\,.
\end{align}

In the angular distribution in Eq. \eqref {ang_distr}, there is also a divergence in the integral over $\tau$  at a frequency
\begin{equation}
\omega=\omega_\theta^*=\dfrac{n\omega^*}{1+u}\,, \quad u=\dfrac{\varepsilon^2\theta_{kp}^2}{m^2(1+\xi^2)}\,,
\end{equation}
where $n\ge 1$ is an integer number. This divergence is also due to the possibility of a cascade process.

In the vicinity of $\omega_\theta^*$ at $\xi^2\ll 1$, we have the following asymptotic form for $n=1$ and $|\omega/\omega^*-1|\ll \xi$:
\begin{align}
&\frac{d\sigma}{d\omega d\bm k_\perp}=\frac{8\alpha\eta^2\xi^2\varepsilon^2b^2\,\omega_\theta^*}{\pi m^4 (1+b^2)^4(\omega-\omega^*_\theta)^4}\left[ L-5/2+\log(b+1/b)+\log|\omega/\omega^*_\theta-1|\right]\,,
\end{align}
where  $b={\varepsilon\theta_{kp}}/{m}$.

Analogously, for $\omega\ll \omega^*_\theta$ and any value of $\xi$, we have 
\begin{align}
\frac{d\sigma}{d\omega d\bm k_\perp}=&\frac{8\eta^2\alpha\varepsilon^2}{\pi m^4(1+u)^4\omega^3}\Bigg\{\left[L-3/2+\log{(1+u)}+\dfrac{1}{2}\log(1+\xi^2)\right](1+u^2)
+2u\Bigg\}\,.
\end{align}

Finally, we note that in all cases there is no divergence for a finite laser pulse, due to the finite duration of the interaction of the combined laser and atomic fields with the electron.

\section{Weak laser field}\label{sec5}
The effects of the laser field depend on the values of the two parameters, $\xi$  and $\chi=(\varepsilon\omega_0/m^2)\xi$ \cite{Ritus_1985,Di_Piazza_2012}. As it was pointed out above, for optical lasers  the relation  $\chi\ll\xi$ holds up to electron energies of the order of $20\,\mbox{GeV}$. For a routinely achievable intensity  of $I=10^{18}\mbox{W/cm}^2$ we have that we have $\xi=0.51$ at $\lambda=850\,\mbox{nm}$ and $\chi=0.015$ for $\varepsilon=5\,\mbox{GeV}$. Since $\mathcal{A}_\tau\propto \xi$,   we can  expand Eq.  \eqref{spectr} with respect to $\mathcal{A}_\tau$  for  $\xi^2\ll 1$ and write $d\sigma/{d\omega}$ as the sum $d\sigma/{d\omega}=d\sigma_a/{d\omega}+d\sigma_l/{d\omega}$, where
\begin{align}\label{spectr_small}
&\dfrac{d\sigma_a}{d\omega}=\frac{4\alpha\eta^2}{m^2\omega}\Bigg[\left(1+\frac{\varepsilon'^2}{\varepsilon^2}-\frac{2\varepsilon'}{3\varepsilon}\right)
L+\frac{1}{9}\frac{\varepsilon'}{\varepsilon}\Bigg]\,,\nonumber\\
&\dfrac{d\sigma_l}{d\omega}=-\frac{4\alpha\eta^2\omega}{m^2D\varepsilon^2}\Re \int\frac{ d \tau_1 d \tau_2}{\tau_+^2} e^{i{\tau_+}}\Bigg\{ \left(2L-C+i\dfrac{\pi}{2}-\log\frac{\tau_1 \tau_2}{\tau_+}\right)\nonumber\\
&\times\left[-\frac{2\tau_1\tau_2\phi}{\tau_+}- (\bm{ \beta}+\bm{ s}_1)\cdot(\bm{ \beta}-\bm{ s}_2)+iD \tau_1\tau_2 (\phi+\zeta) \right]\nonumber\\
&+\frac{3\tau_1\tau_2(\phi+\zeta)}{\tau_+}+ 2\bm{\beta}\cdot(\bm{ s}_1 -\bm{ s}_2)-\bm{ s}_1\cdot\bm{ s}_2-iD \tau_1\tau_2 (\phi+2\zeta) \Bigg\}\,,
\end{align}
with the used notations given in  Eq. \eqref{spectr}. For the case of the circularly-polarized monochromatic plane wave considered above, all integrals can be taken and the result is:
\begin{align}\label{spectr_circ}
&\dfrac{d\sigma_l}{d\omega}=\frac{4\alpha\eta^2\xi^2\omega}{m^2D\varepsilon^2}\Re\Bigg\{
L+\dfrac{1}{2}+\frac{6 L\left(9 y ^2-7\right)-3 y ^2+1}{18\left(y ^2-1\right)^2}+ y^2\bigg[\left(L-\frac{3}{2}\right)l_1\nonumber\\
&-\frac{l_1^2+l_2^2}{8} 
-\frac{2 y  \left(3 y ^2-2\right) l_2+\left(5-7 y ^2\right) l_1}{12 \left(y ^2-1\right)^2}-\frac{1}{2} \text{Li}_2(\frac{1}{y^2})\bigg]\nonumber\\
&-8y^3\frac{\varepsilon'}{\varepsilon}\bigg[\frac{L(y^2-1)}{y^3}+\frac{(12 L+1) \left(7 y ^2-5\right)}{72 y^3 \left(y ^2-1\right)}+\frac{\left(y  l_1-l_2\right)}{12 \left(y ^2-1\right)}\nonumber\\
&-\left(L-\frac{3}{2}\right) \left(y l_1+l_2\right)+(y -1) \text{Li}_2\left(\frac{1}{y }\right)+(y +1) \text{Li}_2\left(-\frac{1}{y }\right)\bigg]\Bigg\}\,, 
\end{align}
where $y=\omega/\omega^*$, $l_1=\log (1-1/y ^2)$, $l_2=\log[ (y +1)/(y -1)]$.

In the vicinity of the resonance (at $|1-y|\ll 1$) we have
\begin{align}\label{spectr_resweak}
&\dfrac{d\sigma_l}{d\omega}=\frac{4\alpha\eta^2\xi^2\omega^*}{3 m^2 (\omega^*-\omega)^2}
\left[L+\frac{1}{2}\log|1-\omega/\omega^*|-\frac{1}{6}\right]\,.
\end{align}
Outside of the resonance we obtain
\begin{align}
\frac{d\sigma_l}{d\omega}=&-\frac{16\eta^2\xi^2\alpha}{3m^2\omega}\left(L-\frac{5}{12} \right)
\end{align}
for $y\ll 1$,  and 
\begin{align}
&\dfrac{d\sigma_l}{d\omega}=\frac{16\alpha\eta^2\chi^2\varepsilon'^2}{m^2\omega^3}\Bigg[
10\left(1+\frac{\varepsilon'^2}{\varepsilon^2}-\frac{64}{75}\frac{\varepsilon'}{\varepsilon}\right)L-9
\left(1+\frac{\varepsilon'^2}{\varepsilon^2}-\frac{1756}{2025}\frac{\varepsilon'}{\varepsilon}\right)
\Bigg]\,
\end{align}
for $y\gg 1$. This asymptotics agrees with the corresponding result in Refs. 
 \cite{BKS88,Baier_b_1998}. 

\begin{figure}[h]
	\centering
	\includegraphics[width=0.6\linewidth]{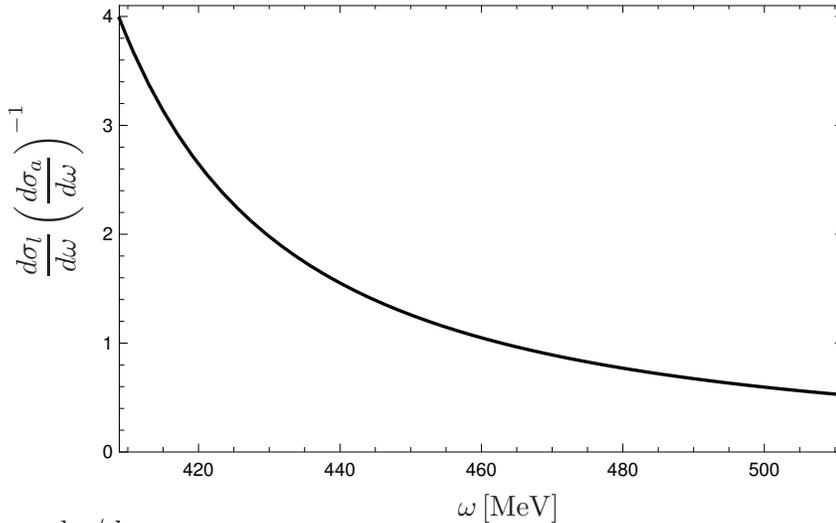}
	\begin{picture}(0,0)(0,0)
	\put(-150,-15){$\omega\,[\mbox{MeV}]$}
	\put(-320,75){\rotatebox{90}{$\dfrac{d\sigma_l}{d\omega}\left(\dfrac{d \sigma_a}{d\omega}\right)^{-1}$}}
	\end{picture}
	\caption{Ratio $\dfrac{d\sigma_l/d\omega}{d \sigma_a/d\omega}$ as a function of $\omega$ at $\xi=0.5$ and $\varepsilon=5\,\mbox{GeV}$ (corresponding to $\chi=0.015$). For these parameters it is $\omega^*=371\,\mbox{MeV}$.}
	\label{pic}
\end{figure}

In the region $\omega>\omega^*$ radiation in a pure laser field is almost absent for the parameters regime under investigation, such that this region is the most appropriate from the experimental point of view in order to observe the impact of the laser field on bremsstrahlung in the atomic field. For $1\gg\omega/\omega^*-1\gg 1/(m r_{scr})^2$, the term $d\sigma_l/d\omega$ becomes of the same order of the term  $d \sigma_a/d\omega$ even at $\xi\ll 1$, which is easily accessible experimentally. This statement is illustrated in Fig.~\ref{pic}, where the ratio $({d\sigma_l/d\omega})/({d \sigma_a/d\omega})$ is shown as a function of $\omega$ at $\xi=0.5$ and $\varepsilon=5\,\mbox{GeV}$ (corresponding to $\chi=0.015$), when $\omega^*=371\,\mbox{MeV}$. It is worth noting that these parameters can be achieved by modern high-power lasers without tightly focusing the laser energy such that the plane-wave approximation is well justified.

The figure clearly shows how the presence of the laser field substantially modifies the bremsstrahlung spectrum.

\section{Conclusion}\label{sec6}

In the present paper, we have investigated in detail the impact of a laser field approximated as a plane wave on the process of photon radiation by an ultra-relativistic electron in an atomic field. We have derived the corresponding angular distribution and the spectrum of the radiated photon. By means of the quasiclassical approximation, the obtained results are exact in the parameters of the laser field and the atomic field. In particular, we have shown that the impact of the laser field is significant even for fairly average values of the laser field parameters, which are routinely obtained in the laboratory nowadays. This makes an experimental observation of the influence of the laser field on bremsstrahlung in the atomic field to be a very feasible task.

\section*{Acknowledgments} 
P. A. K. gratefully acknowledges the Max Planck Institute for Nuclear Physics for the warm hospitality and the financial support during his visit.

\end{document}